\documentstyle[prb,twocolumn,aps,psfig]{revtex}
\tighten
\begin{document}
\draft

\title{Applicability of Fermi golden rule and possibility of 
low-field runaway transport in nitrides}

\author{S. M. Komirenko and K. W. Kim }
\address{Department of Electrical and Computer Engineering,
North Carolina State University, Raleigh, North Carolina 27695-7911}
\author{M. A. Stroscio and M. Dutta}
\address{U.S. Army Research Office, P.O. Box 12211, Research Triangle Park,
North Carolina 27709-2211}

\twocolumn[
\maketitle
\widetext \vspace*{-0.5 in}

\begin{abstract}
\begin{center} \parbox{14cm}
{In order to justify applicability of the standard approach of perturbation 
theory for the description of transport phenomena in wide-band polar 
semiconductors with strong electron-phonon interactions, we have compared 
dependences of energy losses to the lattice on the electron drift velocity 
obtained for different materials in the frameworks of (a) a perturbative 
approach based on calculation of the scattering rates from Fermi's golden 
rule and (b) a non-perturbative approach based on the path-integral formalism 
of Thornber and Feynman.  Our results reveal that despite strong 
electron-phonon coupling in GaN and AlN such that intercollision times become 
of the order of the period of phonon oscillation, standard perturbative 
treatment can still be applied successfully for this type of material. Our 
findings also indicate possibility for unique long-distance runaway transport 
in nitrides which may occur at the pre-threshold electric fields. Polaron 
ground state energy and effective masses are calculated for GaN and AlN as 
well as for GaAs and ${\rm Al_2 O_3}$. }
\end{center}
\end{abstract}


]\narrowtext

\section{INTRODUCTION}
One of the main factors determining electron transport characteristics in 
polar semiconductors is scattering of the electrons by 
polar optical phonons. For relatively weak electron-phonon interactions, when 
scattering events can be considered as independent, use of Fermi's golden 
rule for the calculation of energy-dependent frequencies of electron 
transitions provides an adequate description of experimentally obtained 
velocity-field curves. Upon increasing of the interaction, however, the 
polaronic effects induced by autolocalization of an electron by the inertial 
part of the crystal polarization become more prominent and they determine the 
character of the scattering. Intensification of the electron-phonon interaction 
leads eventually to a situation where the average intercollision time becomes 
less than the duration of a collision. Such a strong coupling, therefore, 
requires proper account for the quantum interference effects and makes the 
problem of electron drift essentially nonlinear. This complicates dramatically 
the theoretical treatment of carrier scattering and field-dependent transport 
since, for the given case, the standard perturbation technique is not applicable. 

To ensure energy conservation for the short-time perturbations, the inverse 
scattering rate, $\tau$, must be large enough to satisfy the inequality 
$\tau \gg \hbar/{\Delta \cal E}$, where $\Delta \cal E$ is the electron 
transition energy. 
This criterion, however, does not allow one to conclude whether or not a wide 
class of materials in which the transition energy may become of the order of 
the linewidth can be described successfully in the framework of standard
perturbation approaches. To such a class of materials belong, in particular, 
nitrides of Al and Ga. These semiconductors have been investigated recently
quite intensively due to a number of their unique properties that can be 
utilized in the current state-of-the-art semiconductor technology. However, 
most attempts to describe scattering processes 
in nitrides have been undertaken assuming the 
validity of Fermi golden rule. 
Moreover, despite the possibility of growing the nitrides of group III in 
zincblende-like structures, their crystal structure at ambient conditions is 
wurtzite-like. For wurtzites, it is generally necessary to account for 
optical anisotropy when considering the carrier-optical-phonon interactions. 
Since optical phonon 
spectra in wurtzites are far more complicated than those of cubic crystals, 
the majority of theoretical results have been obtained by ignoring 
the features of the phonons in optically anisotropic media. Recently, a formalism 
has been developed \cite{us} for evaluation of the scattering rates in bulk 
wurtzite-like semiconductors and heterostructures by taking into account 
peculiarities of the phonon spectra obtained in the framework of macroscopic 
dielectric continuum model. \cite{disp} Due to the anisotropy-induced complexity  
of the problem, Ref. \onlinecite{us} took advantage of the perturbation theory. 
However, as indicated by the previously-discussed considerations, the 
validity of such an approach requires independent confirmation. 

The present paper demonstrates applicability of Fermi's golden rule for 
describing adequately the electron-longitudinal-phonon interaction in polar 
materials by comparing the field-velocity dependences obtained in 
the frameworks of (a) the perturbation theory and (b) the non-perturbative 
path-integral approach of Thornber and Feynman \cite{TF} (TF). A supplemental, 
but very important, result obtained from the present investigation is the 
discovery of the possibility for unique low-field long-distance runaway transport 
in materials characterized by strong electron-phonon interactions. Examples of 
such materials are the nitrides of Al and Ga. 
\section{MODEL}
The most systematic and self-consistent approach for evaluating the
long-range polaron ground-state energy $G$, effective mass $m_0$, and carrier 
energy dissipation for both strong- and weak-coupling limits of electron-phonon 
interaction has been developed by Feynman et al. \cite{TF,F,Fetal} 
In Ref. \onlinecite{TF}, the problem of electron drift in a parabolic band under 
steady state conditions is considered quantum mechanically \cite{S} 
assuming that all the 
energy losses are due to interaction of electrons with polar optical modes. 
Taking advantage of the Fr\"{o}hlich's polaron model, the authors used the
path-integral method to eliminate the lattice coordinates from the momentum 
balance equation and obtained an expression for the magnitude of the electric 
field $E$ that is required to maintain a particular magnitude of electron 
velocity $V$ at arbitrary temperature and interaction strength characterized 
by the coupling constant 

\begin{equation}
\alpha=\frac {e^2}{\hbar}\left (\frac {1}{\epsilon_{\infty}}-
\frac {1}{\epsilon_0} \right ) 
\left [ \frac {m^*}{2 \hbar \Omega} \right ]^{1/2}.
\end{equation}
Herein, $e$ is the elementary charge. 
Evaluation of $\alpha$ requires four parameters: electron effective mass $m^*$; 
frequency of the longitudinal phonon $\Omega$; static dielectric constant 
$\epsilon_0$; and high frequency dielectric constant $\epsilon_{\infty}$.  
All these parameters can be measured experimentally and they are the only 
external parameters required for calculation of energy loss per unit 
distance $eE$ versus $V$. 

Thornber and Feynman \cite{TF} have calculated the dependences of 
$eE(V)$ for three coupling constants ($\alpha=3,5,7$) over a wide range of 
reciprocal temperatures 
$\beta=\hbar\Omega/(k_B T)$, where $k_B$ is the Boltzmann constant and $T$ is 
the temperature of the lattice. The general result of these calculations can 
be summarized briefly as follows. For each particular $\alpha$, $eE(V)$ has a 
maximum at some threshold value $V_{th}$. For 
$\beta > 1$, location of this maximum becomes independent of temperature. 
For $V<V_{th}$, $eE$ is an increasing function of velocity. This interval of 
velocities corresponds to a stable situation when energy loss to the lattice 
due to the absorption and emission of optical phonons can 
be compensated by the energy gained by the electron from the applied field in 
such a way that at the given $E$, a small deviation $\Delta V$ of the 
velocity from its steady state value $V_s$ creates a force,  
$e[E(V_s)-E(V_s \pm \Delta V)]$, which stabilizes the velocity at $V_s$. 
When the external field approaches the value $E_{th}=E(V_{th})$, the dependence 
tends to saturate since the magnitude of the energy loss due to interaction with 
optical phonons is finite. The case $E>E_{th}$ was excluded from 
consideration because no steady state conditions can be 
reached for such fields and electron would accelerate infinitely. The theory, 
however, predicts the 
existence of solutions for $V > V_{th}$. In this region, $eE$ is a 
decreasing function of the velocity which leads to an unstable steady state 
situation. For this case, any deviation of the velocity from $V_s$ would lead to 
either deceleration of the electron to velocity $V < V_{th}$ which is stable at 
the given field, or a {\em gradually} 
increasing acceleration if $\Delta V $ leads to an increase in velocity. 
It is essential, that for $V > V_{th}$, 
the dependence $eE(V)$ can be interpreted as a time-dependent momentum loss in 
the absence of the external field. This loss would coincide with the rate of 
electron momentum loss if the criterion $dV/dt \ll V/$(duration of the collision)  
is satisfied. For $\beta > 1$, $eE(V>V_{th})$ also becomes independent on 
temperature.  

	In order to simplify the comparison, and taking into account that the 
strongest electron-polar-optical-phonon scattering is due to emission of the 
longitudinal optical (LO) phonons, we will consider the case when $\beta \approx 4$. 
Due to high energy of LO phonons in the nitrides, such value of $\beta$ would 
correspond to room temperature in these materials. For GaAs, which we take as a 
reference point in our investigation, $\beta =4$ would correspond to lattice 
temperature of order of 104 K. Since $\beta_{GaAs}$ is slightly higher than 1 at the 
room temperature, the result obtained for maximum energy loss per unit distance can 
be compared to the experimental {\em velocity-field} dependences (see, for example, 
Ref.\onlinecite{shur} and citations therein). 
Indeed, at some threshold field $E_{th}$, the dependence $V(E)$ has a maximum caused 
by transitions of the carriers to an upper valley with a higher effective mass. In 
terms of the TF model, these transitions would start to occur when the energy 
supply from the external field would exceed the maximum loss to the lattice; i.e.,  
at $E > E_{th}$. 

Thus, if the average kinetic energy of electrons obtained in the framework of TF 
model at $E_{th}$ does not exceed the energy of bands separation and the effects 
of the increased effective mass due to non-parabolicity of $\Gamma$ band can be 
neglected, the value of the argument at maximum of $V(E)$ dependence has to correlate 
with the extremum of the $eE(V)$. 

	Under the assumptions made above, we use the simplest model for estimation 
of energy loss to the lattice in the framework of the perturbation theory. In this
model, we calculate the dependence of scattering rate $1/\tau$ due to the emission 
of LO phonons on a single electron kinetic energy ${\cal E} $  using the Fermi's 
golden rule. For an optically isotropic material, the scattering rate is
 \begin{eqnarray}
\frac {1}{\tau} = \left ( \frac {2 m^*}{{\cal E}}  \right )^{1/2} 
\frac {e^2 \Omega (N_q+1)}{\hbar} \left ( \frac {1}{\epsilon_{\infty}}-
\frac {1}{\epsilon_0} \right ) \nonumber \\
 \ln \left [\sqrt{\frac {{\cal E}}{\hbar \Omega}-1}+ 
\sqrt{\frac {{\cal E}}{\hbar \Omega}}  \right ],
\end{eqnarray}
where $N_q$ is the phonon occupation number. Then, we assume that a carrier 
with velocity 
$V=\sqrt{2 {\cal E}/m^*}$ loses the energy $\hbar \Omega$ to the lattice in 
a distance  $\tau V$. 

\section{RESULTS AND DISCUSSION}

\subsection{Limiting cases}

The energy loss per unit distance vs. the electron velocity calculated 
for GaAs at room temperature in the framework of the model which uses 
Fermi's golden rule is shown on Fig. \ref{Fig1} by the thin 
solid line. It is important, that as anticipated, the maximum of the dependence 
is in a good agreement with the experimental data which give the maximum of 
$V(E)$ at $E \approx 3.4 - 3.9$ kV/cm. One should note, that since calculations 
are made in a one-electron approximation and for single parabolic band, they 
overestimate the velocity obtained at the maximum. The experimentally measured 
values at maximum of $V(E)$ \cite{shur} reflect the averaging of the 
velocities over bands with different effective masses as well as effects of 
non-parabolicity. 
\begin{figure}
\narrowtext
\hspace{0.0cm}\psfig{figure=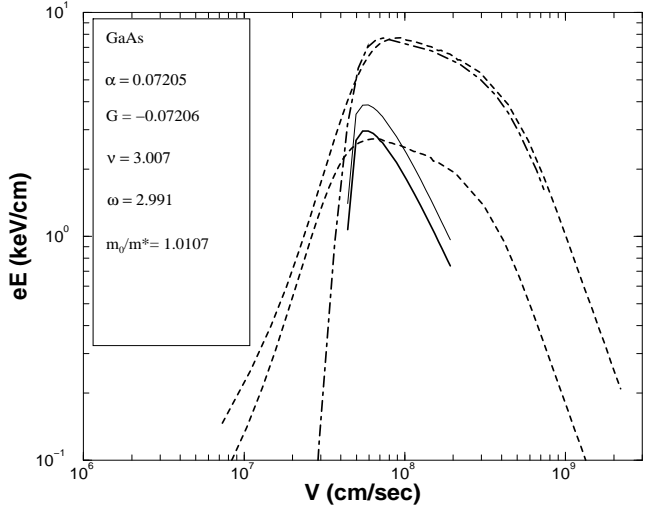,width=10.0cm,height=8.0cm,angle=270}
\protect \vspace{0.1cm}
\caption{The energy loss per unit distance vs. electron velocity in GaAs. 
Perturbative model: room temperature - thin solid line; T=104 K - thick solid line. 
TF model: T=104 K - upper dashed line; T $\approx$ 20 K - dash-dotted line. 
Corrected TF model: T=104 K - lower dashed line. Polaron parameters are 
given in the inset.}
\label{Fig1}\end{figure}
The curve calculated in the simplest model at 104 K ($\beta=4$) is given by the 
thick solid line. Surprisingly, the $eE(V)$ dependence computed for $\beta=4$ in the 
framework of the TF model (upper dashed line) exhibits a maximum located at somewhat 
higher fields. We assume that 
such a discrepancy occurs because in a weak-coupling limit the 
zero-order distribution for the electrons in this model reduces to a drifted 
quasi-Maxwellian. An essential requirement for such a distribution to be valid in the 
given case is the presence of high electron concentration and strong electron-electron 
interactions, which provide randomization of the direction of electron momentum 
between the scattering events. \cite{frol} It is unlikely, however, that such a 
randomization can be achieved for scattering with emission of polar optical phonons. 
Indeed, in the weak-coupling limit the motion of the carrier in the near-threshold 
fields becomes essentially one-dimensional. Conservation of energy and momenta, valid 
in the weak-coupling limit, require the angle between the carrier and scattered 
phonon momenta to be no more than $\arccos(\hbar \Omega/{\cal E})$. Due to this 
condition, emission of an LO phonon at threshold fields causes deviation of the 
direction of electron momentum from the direction of applied field by no more than 
$\approx$ 20 degrees. Additional focusing of the electron momentum in the direction 
of the electric field comes from the inverse dependence of the interaction matrix 
element on the phonon wavevector \cite{dumke} and leads to overestimation of energy 
losses when using the Maxwellian distribution. \cite{KH}  

	In order to resolve this discrepancy, one can suggest - in analogy to the 
classical case - that at the same electron temperature, reduction of the dimension 
would correspond to a reduction in the average carrier energy. Since the deviation 
of the direction of the electron  momentum relative to the direction of the 
electric field is small but finite, for the weak-coupling limit we have 
reduced the energy scale in TF model by a factor of two in order to match the maxima. 
This  corresponds to decreasing the velocity and energy losses by factors $2^{1/2}$ 
and $2^{3/2}$, respectively. The corrected curve $eE(V)$ is shown by lower dashed 
line. 

The differences in the shapes of the curves obtained in the simplest model and the 
TF model occur for the following reasons. The former model considers only the emission 
mechanism, whereas the latter model also takes into account absorption. As shown in 
the figure by dash-dotted line calculated for the uncorrected case of TF model for 
$\beta \approx 21$, elimination of phonons absorption by reducing the lattice temperature 
yields the same slope of $eE(V < V_{th})$ as in the simple model. As expected, for 
this temperature interval, the temperature decrease does not affect the shape of the 
curve at and beyond the  maximum. For $V > V_{th}$, the discrepancy in $eE(V)$ between 
the models appears to be due to the relatively high value of $d V/dt$ in the unstable 
region. In the framework of the TF model, one can estimate this value by $eE /m_0$, 
where $m_0$ can be obtained as $m_0 \approx m^* \nu^2/\omega^2$; $\nu$ and $\omega$ 
are the parameters of TF model.  We have computed these parameters from minimization 
of the free energy at zero temperature. \cite{F} Our estimations show that assuming 
the duration of the collision to be equal to $\tau$, the value of the derivative would 
be much less than $V/\tau$ only for $V \sim 10^9$ cm/sec. Thus, for $V > V_{th}$ the 
simple model cannot be used for GaAs under the conditions when an external field is 
applied. 

	Additional confirmation of the idea that the corrections required for the 
application of the TF model in the weak-coupling limit are induced by focusing of the 
carrier momentum comes from the fact, that in the case of strong coupling - when the 
directions of momenta are randomized due to strong electron-phonon interaction 
discussed previously - the model correctly explains the experimentally-obtained results. 
In Fig. \ref{Fig2} we depict the energy losses calculated for ${\rm Al_2O_3}$. The 
maximum on the dependence obtained in the TF model (dashed curve) is in the excellent 
agreement with the experimentally obtained maximum losses in this material, 
0.03 eV/$\AA$. \cite{Handy}
\begin{figure}
\narrowtext
\hspace{0.0in}\psfig{figure=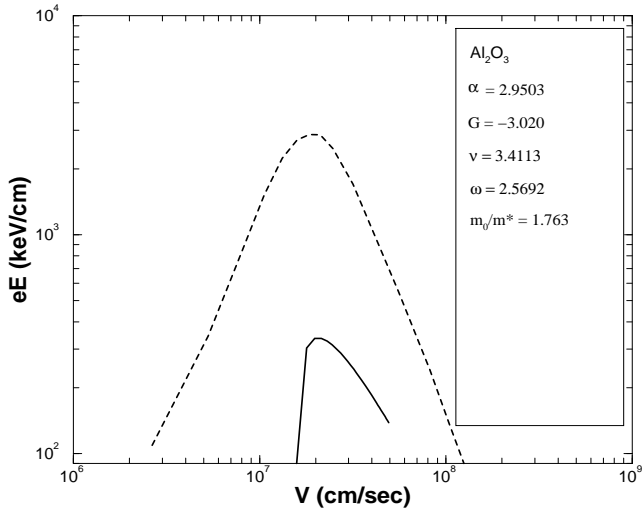,width=10.0cm,height=8.0cm,angle=270}
\protect \vspace{0.1cm}
\caption{The energy loss per unit distance vs. electron velocity in ${\rm Al_2O_3}$ at 
room temperature. Perturbative model: solid line; TF model: dashed line; Polaron 
parameters are given in the inset.}
\label{Fig2}\end{figure} 
As expected, the maximum losses calculated in the perturbative model (solid line) are 
less by an order of magnitude. Of course, since for the given material $\alpha > 1$, 
the simple model is not valid and we have presented here both dependences simply as a 
means of estimating the possible error which can be induced by a perturbative 
treatment and to demonstrate that no energy scale reduction can fit these dependences. 
\subsection{Intermediate case: nitrides}

	Due to the optical anisotropy inherent to wurtzites, the coupling parameter of 
polaron theory $\alpha$  becomes dependent on the angle $\theta$ between 
phonon wavevector and the optical axis.   Assuming that 
$\epsilon_z^{\infty}=\epsilon_t^{\infty}$, we define this dependence as 

\

\begin{eqnarray}
\alpha(\theta)=\frac {e^2}{\epsilon^{\infty} \Omega} \sqrt{\frac {m^*}
{2(\hbar \Omega)^3}} 
\left [  \frac {\omega_{Lz}^2-\omega_z^2} {(\Omega^2-\omega_z^2)^2}
\cos^2\theta \right. \nonumber \\
\left.+\frac {\omega_{Lt}^2-\omega_t^2} 
{(\Omega^2-\omega_t^2)^2} \sin^2 \theta
 \right ]^{-1},
\end{eqnarray}
where $\omega_{Lz}$, $\omega_{z}$, $\omega_{Lt}$, and $\omega_{t}$ 
are the characteristic frequencies of the 
$\rm A1(LO)$, $\rm A1(TO)$, $\rm E1(LO)$, and $\rm E1(TO)$ modes, 
respectively.  The phonon frequency as a function of $\theta$ can be obtained from the 
dispersion relation for the extraordinary bulk phonons. \cite{disp}
The dependence $\alpha(\theta)$ calculated for GaN  is shown on Fig. \ref{Fig3}. 
\begin{figure}
\narrowtext
\hspace{0.0in}\psfig{figure=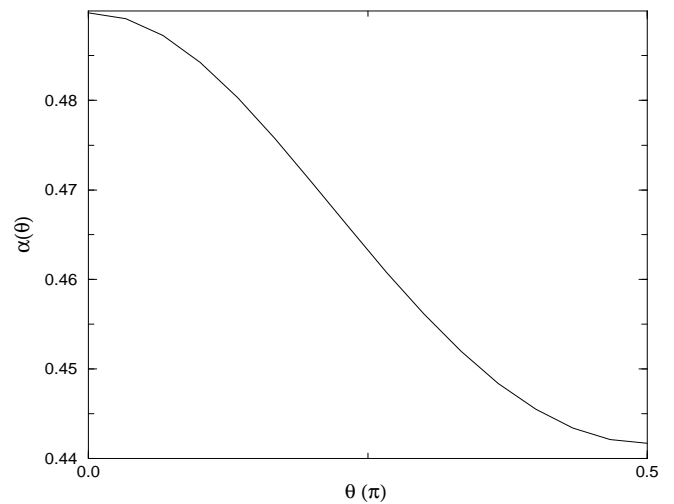,width=10.0cm,height=8.0cm,angle=270}
\protect \vspace{0.1cm}
\caption{Dependence of the polaron coupling parameter on the angle 
between the phonon wavevector and the optical axis in GaN.}
\label{Fig3}\end{figure}
To obtain the $eE(V)$ dependence in the TF model we have used $\alpha = 0.46$. 
This value corresponds to the 
energy of LO phonon calculated for GaN in the cubic phase. \cite{Karch} The energy 
of the LO phonon in cubic AlN is taken to be 113 meV which is - as for the GaN case -  
between the energies of A1 and E1 LO modes in wurtzite phase. \cite{disp} 
The scattering rates are calculated according to the formalism developed in 
Ref. \onlinecite{us}. 
Comparison of the dependences obtained for the nitrides is given in Fig. \ref{Fig4}.  
Again, the maxima obtained in the simple model correlate very well with the threshold 
electric fields of $V(E)$ dependences computed in the Monte Carlo technique \cite{foutz}
for a three-valley model for the conduction band: 
140 kV/cm for GaN and 450 kV/cm for AlN. 
Note that in order to match the maxima one needs to use the same reduction of the 
energy scale when calculating the energy losses in the TF model as for the case of GaAs. 
Additionally, one can see that the shapes of the dependences for $V \geq V_{th}$ 
are almost the same. This agreement between the simple and the TF models is due to 
the extremely short duration of the collisions which can be estimated  
roughly as the inverted scattering rate, $\tau \sim 10^{-14}$ sec. The increase of the 
polaron effective mass cannot compensate the increase in the energy loss 
near the threshold value and, therefore, cannot reduce dV/dt. Nevertheless, due to 
frequent collisions, the criterion $dV/dt \ll V/\tau$ is satisfied for the nitrides 
even at $V \rightarrow V_{th}^+$. 
\begin{figure}
\narrowtext
\hspace{0.0in}\psfig{figure=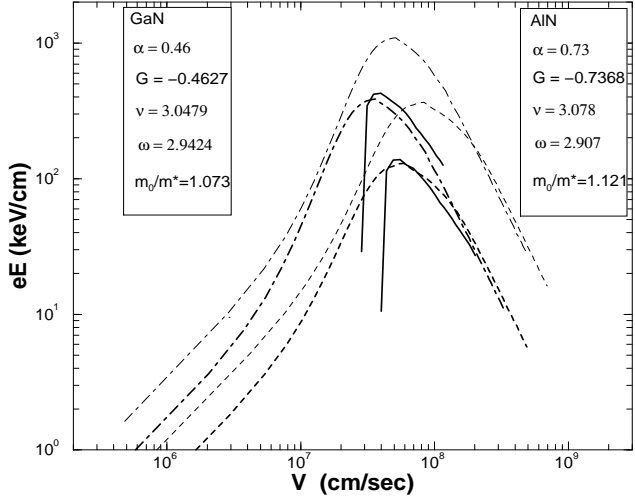,width=10.0cm,height=8.0cm,angle=270}
\protect \vspace{0.1cm}
\caption{The energy loss per unit distance vs. electron velocity in GaN and AlN calculated 
at the room temperature. 
Perturbative model: AlN - upper solid line, GaN - lower solid line; 
TF model: AlN - dash-dotted lines, GaN - dashed lines. Thin and thick broken curves are 
the uncorrected and corrected dependencies, respectively.
Polaron parameters are given in the insets.}
\label{Fig4}\end{figure}
The comparison made here allowed us to conclude that in materials for 
which $\tau \Omega \gtrsim 1$,  application of Fermi golden rule to explain transport 
phenomena is as good as in materials traditionally handled with the perturbation theory, 
i.e., in the materials for which much stronger criterion, $\tau \Omega \gg 1$, is 
satisfied.

\subsection{Low-field runaway transport}

	It is important to emphasize another result of our 
investigation. Figure \ref{Fig5} 
represents the dependencies of energy losses on electron kinetic energy obtained in the 
framework of the TF model at room temperature for the three  materials considered: 
GaAs, GaN and AlN. On the figure, vertical arrows indicate the energy of the closest 
upper valley in the corresponding conduction band. 
The picture shows clearly the possibility to achieve unique pre - threshold - field 
runaway transport in the nitrides. Indeed, let us consider a GaN sample in an external 
electric field of 100 kV/cm. As shown in the figure, the two steady-state 
solutions  for electron energy would correspond to such a situation. One of the solutions 
lies in the region $V < V_{th}$ and, consequently, it reflects a stable solution with 
respect to the electron energy fluctuations. Another one falls into the unstable area, 
$V > V_{th}$. Suppose an electron is injected into the sample with an energy somewhere 
in between the energy which corresponds to the second, unstable, solution and the 
threshold energy. 
\begin{figure}
\narrowtext
\hspace{0.0in}\psfig{figure=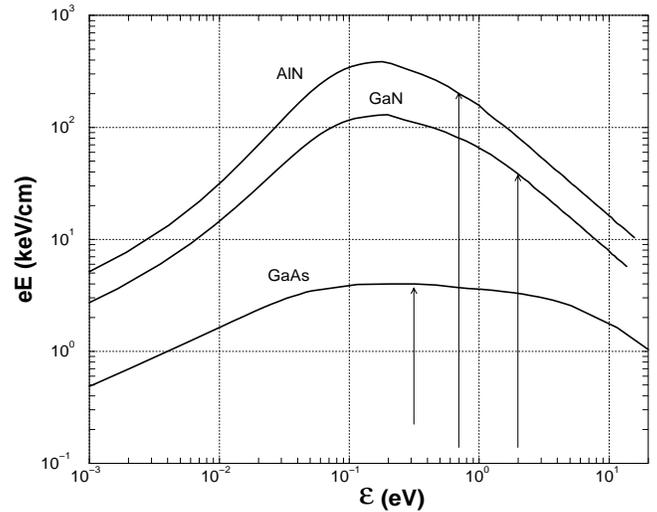,width=10.0cm,height=8.0cm,angle=270}
\protect \vspace{0.1cm}
\caption{The energy loss per unit distance vs. electron kinetic energy. Vertical arrows indicate the 
energy of the closest upper valley in the corresponding conduction band. T = 300 K.}
\label{Fig5}\end{figure}
In this case, since the energy losses to the lattice would 
exceed the energy gain from the external field, the electron would decelerate until the 
stable solution at given field would be reached. If, however, 
the energy of the injected electron would just slightly exceed the value of high-energy 
steady-state solution, the electron would accelerate moving downwards on the unstable 
branch of the dependence until it gains enough energy, ${\cal E}^*$, to appear in the 
nearest upper valley of the conduction band. The value of acceleration would gradually 
increase due to the increasing difference between the force caused by the external 
field, $F = 100$ keV/cm, and the force caused by energy losses to the lattice 
represented by the $eE({\cal E})$ dependence.  In order to estimate the minimum 
runaway length $L_r$ , we have assumed that the energy of the injected electron is 
0.7 eV and that the maximum force, $F-eE({\cal E}^*)$ , is constantly applied to the 
carrier. Under this assumption, we obtained  $L_r^{GaN} > 220$ nm. 

It is easy to see that because the intervalley separation in AlN is smaller, the effect 
in this material is expected to be not as strong as in the previous case. Assuming 
injection energy 0.31 eV and applying field 300 kV/cm, we get $L_r^{AlN}>39$ nm. Our 
results also show that the previously-discussed runaway transport cannot be achieved 
in GaAs due to the small intervalley gap and broad peak on the $eE({\cal E})$ dependence.

The results presented in this paper were obtained ignoring the non-parabolicity effects. 
These effects, however, would not change our main findings qualitatively. In order 
to improve accuracy of the expected quantities, further investigation is required. 
It is interesting to note, that the abstract possibility of the low-field runaway 
transport was mentioned initially by Thornber and Feynman. \cite{TF} Nitrides of Ga 
and Al promise to be materials where such transport could be actually realized. 

\section{SUMMARY}

In present paper, we have compared the energy losses to the lattice calculated in 
different polar semiconductors within the frameworks of both non-perturbative and 
perturbative approaches. Our results reveal that application of Fermi golden rule for 
calculation of the scattering rates in nitrides, where $\tau \Omega \approx 1.5$,  is 
as appropriate as application of this standard perturbative treatment in the materials 
for which the well-known criterion $\tau \Omega \gg 1$ is satisfied. This finding 
dramatically simplifies the analysis of transport phenomena in the wide-band polar 
semiconductors with intermediate magnitude of polaron coupling factor, $\alpha < 1$.  

Applying the non-perturbative path-integral approach of Thornber and Feynman to 
evaluation of field dependent electron energy dissipation in AlN, GaN and GaAs, we 
have found that pre-threshold low-field runaway electron transport can be realized in 
the nitrides. The conditions for such a transport can be formulated as follows: 
(a) the energy of the injected carrier should exceed the energy which corresponds to 
the solution of the momentum balance equation located on the unstable branch of $eE(V)$; 
and (b) the separation  between the energy of injected carrier and the energy of 
the bottom of an upper valley  must be high enough to provide a finite value of 
runaway length. It must be at least a few times higher than the energy of the polar 
optical phonon.

\section{Acknowledgments}
This study was supported, in part, by the 
Office of Naval Research and by the U.S. Army Research Office.


\begin{references}

\bibitem{us} S. M. Komirenko, K. W. Kim, M. A. Stroscio, and M. Dutta, 
Phys. Rev. B  {\bf 61}, 2034 (2000).

\bibitem{disp} S. M. Komirenko, K. W. Kim, M. A. Stroscio, and 
M. Dutta, Phys. Rev. B {\bf 59}, 5013 (1999).

\bibitem{TF} K. K. Thornber and Richard P. Feynman, Phys. Rev. B {\bf 1}, 4099 
(1970).

\bibitem{F} R. P. Feynman, Phys. Rev. {\bf 97}, 660 (1955).

\bibitem{Fetal} R. P. Feynman, R. W. Hellwarth, C. K. Iddings, and P. M. Platzman, 
Phys. Rev. {\bf 127}, 1004 (1962).   

\bibitem{S} M. A. Stroscio, in {\it Introduction to Semiconductor Technology:
GaAs and Related Compounds}, edited by Cheng T. Wang (Wiley, New York, 1990).

\bibitem{shur} Michael Shur, {\em Physics of Semiconductor Devices} 
(Prentice, New Jersey, 1990).

\bibitem{frol} H. Fr\"{o}hlich and B. V. Paranjape, Proc. Phys. Soc. (London) {\bf B69}, 21 
(1956). 

\bibitem{dumke} W. P. Dumke, Phys. Rev.  {\bf 167}, 783 (1967).

\bibitem{KH} K. Hess, in {\it Physics of Nonlinear Transport in Semiconductors},
edited by D. K. Ferry, J. R. Barker, and C. Jacobiani (Plenum, New York, 1980).

\bibitem{Handy} R. M. Handy, J. Appl. Phys. {\bf 37}, 4620 (1966).   

\bibitem{Karch} K. Karch, F. Bechstedt, and T. Pletl, Phys. Rev. B {\bf 56}, 3560 
(1997).

\bibitem{foutz} B. E. Foutz, S. K.  O'Leary, M. S. Shur, 
and L. F. Eastman, J. Appl. Phys. {\bf 85}, 7727 (1999). 

\end{references}
\end{document}